\documentclass[
 aip,
 jmp,
 amsmath,amssymb,
 reprint,
]{revtex4-1}

\usepackage{graphicx}
\usepackage{dcolumn}
\usepackage{bm}

\begin{document}
\preprint{}
\title{A general method for identifying node spreading influence via the adjacent matrix and spreading rate}
\author{Jian-Hong Lin}
\affiliation{Research Center of Complex Systems Science, University of Shanghai for Science and Technology, Shanghai 200093, P. R. China}
\author{Jian-Guo Liu}
\email[]{liujg004@ustc.edu.cn}
\affiliation{Research Center of Complex Systems Science, University of Shanghai for Science and Technology, Shanghai 200093, P. R. China}
\author{Qiang Guo}
\affiliation{Research Center of Complex Systems Science, University of Shanghai for Science and Technology, Shanghai 200093, P. R. China}
\date{\today}
\begin{abstract}
With great theoretical and practical significance, identifying the node spreading influence of complex network is one of the most promising domains. So far, various topology-based centrality measures have been proposed to identify the node spreading influence in a network. However, the node spreading influence is a result of the interplay between the network topology structure and spreading dynamics. In this paper, we build up the systematic method by combining the network structure and spreading dynamics to identify the node spreading influence. By combining the adjacent matrix $A$ and spreading parameter $\beta$, we theoretical give the node spreading influence with the eigenvector of the largest eigenvalue. Comparing with the Susceptible-Infected-Recovered (SIR) model epidemic results for four real networks, our method could identify the node spreading influence more accurately than the ones generated by the degree, K-shell and eigenvector centrality. This work may provide a systematic method for identifying node spreading influence.
\end{abstract}
\pacs{89.20.Hh, 89.75.Hc, 05.70.Ln}
\maketitle
\section{Introduction}
Spreading is a widespread process in nature, which describes many important activities in society \cite{Kitsak201001,Klemm201202,Castellano201203,Zhou200604}, such as the virus spreading \cite{Kephart1997}, reaction diffusion process \cite{Colizza2007,Liu2007}, pandemics \cite{Pastor2001}, cascading failures \cite{Motter2004} and so on. The knowledge of the spreading pathways through the network of interactions is important for developing effective methods to either hinder the disease spreading, or accelerate the information dissemination spreading. So far, there are a lot of works focusing on identifying the node spreading influence in a network \cite{Ghoshal20111,Borge20122,Borge20123,Liu201303,Zeng2013,Ren201302,Ren201303,Ren 201401,Orsini2013}. Related classical centrality methods include the degree as the number of the node's neighbors, eigenvector centrality \cite{Borgatti2005} as the eigenvector of the largest eigenvalue of the adjacent matrix, K-shell centrality \cite{Kitsak201001} as an effective algorithms based on node location that outperform the classical centrality methods£¬ the closeness centrality \cite{Sabidussi1966} as the reciprocal of the sum of the geodesic distances to all other nodes, betweenness centrality \cite{Freeman1977,Freeman1979} as the number of shortest paths through a certain node. Lately, a lot of works tried to improve the classical methods and proposed effective methods for identifying node spreading influence. For example, Sabidussi\cite{Sabidussi1966} and Chen {\textit{et al}} \cite{Chen2012,Chen2013,Dangalchev2006,Zhang201101} focused on directly improving the basic centrality measures including degree, closeness and betweenness. Liu and Zeng \cite{Liu201303,Zeng2013} tried to improve the K-shell method by removing the degeneracy of the method. Poulin \cite{Poulin2000} focused to cut down the computational complexity of the eigenvector. Moreover, the concept of path  diversity is used to improve the ranking of spreaders \cite{Chen201301}. Liu and Ren \cite{L201101,Ren201303} also designed in directed networks to identify the influential spreaders such as LeaderRank, which is shown to outperform the well-known PageRank method in both effectiveness and robustness.

The above classic and improved centrality methods are based on the network topology structure. However, the node spreading influence is determined not only by the network structure but also by the spreading dynamics \cite{Liu201302,Borge-Holthoefer201201,Borge-Holthoefer201202,Klemm20212,Aral2012,Liu2007}. The study of spreading dynamics is a promising domains that is finding more and more applications in a wide range of areas and it also can help us to understand the unfold of dynamical processes in complex networks \cite{Pastor2014}.  Therefore it is necessary to build up the systematic method to identify the node spreading influence by combining the network structure and spreading dynamics. In this paper, we design a structure spreading dynamics (SSD) method for identifying node spreading influence. Since the adjacent matrix can reflect the network structure, we build up a differential equation by the network adjacent matrix and the spreading process. Then the node spreading influence under different time step $t$, spreading rate $\beta$ and recovering rate $\mu$ can be identified by function of adjacent matrix $A$. To evaluate the performance of the SSD method, the Kendall's tau $\tau$ is introduced to measure the correlation between the ranking list from different centralities and the ranking list from the true spreading influence. The results show that the SSD method can identify the node spreading influence centrality methods. This work provides a systematic method for ranking the node spreading influence.
\section{method}
In this section we will introduce some basic connect from graph theory which will be used in the rest of paper.

\begin{figure}[ht]
\center\scalebox{0.385}[0.385]{\includegraphics{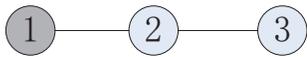}}
\caption{An example network consisted 3 nodes and 2 edges. Node 1 is an initial infected node. It would infect its neighbour node 2 with probability $\beta$ and recover with probability $\mu$ at time step 1.}\label{Fig1}
\end{figure}

Normally, An undirect network $G=(N,E)$ with $N$ nodes and $E$ edges could be described by an adjacent matrix $A=\left\{a_{ij}\right\}$
where $a_{ij}=1$ if node $i$ is connected by node $j$, and $a_{ij}=0$ otherwise. For an undirect network, $A$ is binary and symmetric with zeros along the main diagonal. Therefore, the eigenvalues of $A$ will be real. We label the eigenvalues of $A$ in descending order: $\lambda_{1}{\geq}\lambda_{2}{\geq}{\ldots}{\geq}{\lambda}_{n}$. Since $A$ is a symmetric and real-valued matrix, $A=Q{\Lambda}Q^{T}$, where $\Lambda=diag(\lambda_{1},\lambda_{2}{\ldots},{\lambda}_{n})$, $Q=[\textbf q_{1},\textbf{q}_{2},\ldots,\textbf{q}_{n}]$ and $\textbf{q}_{i}$ is the eigenvector of eigenvalue of $\lambda_{i}$.

Implementing the SIR \cite{Kitsak201001} spreading process for one network, in the SIR model, There are three compartments: (i) Susceptible individuals represent the individuals (not yet infected) who are easy to be infected; (ii) Infected individuals represent individuals who have been infected and are able to spread the disease to susceptible individuals; (iii) Recovered individuals represent individuals who have been recovered and will never be infected again. In each time step, we denote that all nodes are initially susceptible except only one infectious node. The infected nodes will infect their susceptible neighbors with the spreading rate $\beta$, and infected nodes would recover with recovering rate $\mu$ in the next time step. The number of infections generated by the initially-infected node is denoted as its spreading influence. For each initial node, the node spreading influence is obtained by averaging over 100 independent runs and 10 time steps in Fig. 2-3.

We now introduce the structure spreading dynamics (SSD) method. We build up a systematic method by differential equation by combining the adjacent matrix $A$ spreading process. The basic idea is that an infected node would infect its neighbours with spreading rate $\beta$ and recover or remove with the recovering rate $\mu$. We denoted $x_{i}(t)$ is the state of node $i$ at time step $t$. $\textbf{x}(0)$ is the initial state of a network. If $\textbf{x}_{i}(0)=1$ and $\textbf{x}_{j\neq{i}}(0)=0$, node $i$ is initial infected node. Therefore, $\textbf{x}(t)-\textbf{x}(t-1)$ is the probability of the nodes to be infected at time step $t$. We can approximate by the linearization
\begin{equation}\label{equation1}
\textbf{x}(t)-\textbf{x}(t-1)={\beta}{A}[{\beta}{A}+(1-\mu)I]^{t-1}{\textbf{x}(0)},
\end{equation}
where $\beta$ is the spreading rate, $\mu$ is the recovering rate, $A$ is the network adjacent matrix , $I$ is a $N\times{N}$ unit matrix and $\textbf{x}(0)$ is the initial state of network. As shown in Fig. 1, node $1$ is an initial infected node. Therefore, $\textbf{x}(0)=[1,0,0]^{T}$ and the probability of the nodes to be infected at time step $1$ would be $\textbf{x}(1)-\textbf{x}(0)=\beta{A}\textbf{x}(0)=[0,\beta,0]^{T}$. The total probability $\textbf{x}(t)-\textbf{x}(0)$ of the nodes to be infected at time step $t$ would be
\begin{equation}\label{equation2}
\begin{aligned}
 \textbf{x}(t)-\textbf{x}(0)=\sum_{k=1}^{t}[\textbf{x}(k)-\textbf{x}(k-1)] \\
 {=\sum_{k=0}^{t-1}{{\beta}A[{\beta}A+(1-\mu)I]^{k}\textbf{x(0)}}.}\\
 \end{aligned}
\end{equation}
The node spreading influence of node $i$, $\textbf{S}_{i}(t)$, could be appoximate calculated by the following way
\begin{equation}\label{equation3}
\textbf{S}_{i}(t)=\{\sum_{k=0}^{t-1}{{\beta}A[{\beta}A+(1-\mu)I]^{k}\}^{T}\textbf{l}_{i}},
\end{equation}
where $\textbf{l}$ is a $N\times1$ matrix whose components are 1. When recovering rate $\mu=0$ and $\mu=1$, $\textbf{S}_{i}(t)$ is the spreading influence of node $i$ for SI and standard SIR model at time step $t$ respectively.

The spreading influence of node $i$, $\textbf{S}_{i}(t)$, can be written in the following way by decomposing the adjacent matrix $A$,
\begin{equation}\label{equation4}
\textbf{S}_{i}(t)=m_{1}\textbf{q}_{1i}\sum_{j=1}^{n}\textbf{q}_{1j}+\sum_{k=2}^{n}m_{k}\textbf{q}_{ki}\sum_{j=1}^{n}\textbf{q}_{1j},
\end{equation}
where $m_{k}=(\mu-\beta{\lambda_{1}})\{\beta{\lambda_{k}}[1-(\beta{\lambda_{k}+1-\mu})]\}^{-1}$. Let $\varphi_{i}(t)=(m_{1}\sum_{j=1}^{n}\textbf{q}_{1j})^{-1}\textbf{S}_{i}(t)$, Then
\begin{equation}\label{equation5}
\varphi_{i}(t)=\textbf{q}_{1i}+(m_{1}\sum_{j=1}^{n}\textbf{q}_{1j})^{-1}\sum_{k=2}^{n}m_{k}\textbf{q}_{ki}\sum_{j=1}^{n}\textbf{q}_{1j},
\end{equation}

The ranking list generated by $\varphi(t)$ is the same as $\textbf{S}(t)$. Since $\lambda_{1}>\lambda_{k}$, for $2\leq{k}\leq{n}$, as $\beta\rightarrow{1}$ and $t\rightarrow{\infty}$ we can find that $\varphi(t)\rightarrow{\textbf{q}_{1}}$. By the Perron-Frobenius Theorem \cite{Hom1985} $\textbf{q}_{1}>0$. Thus when $\beta\rightarrow{1}$ and $t\rightarrow{\infty}$, the ranking list generated by SSD method is the same as the one generated eigenvector centrality.

\section{Experiment Results}
\subsection{Data description}
To check the performance of the SSD method, two real networks are introduced in this paper including the Email \cite{Guimera2003} and Protein networks. The Email network of University Rovira i Virgili (URV) of Spain contains faculty, researchers, technicians, managers, administrators, and graduate students. The Protein network is a protein-protein interaction network in budding yeast.

The statistical properties of two real networks are shown in Table \uppercase\expandafter{\romannumeral1}, including the number of nodes $N$, edges $E$, the average degree $\langle k\rangle$ and the largest eigenvalue $\lambda_{max}$. 
\begin{table}
\caption{Basic statistical features of Email and Protein networks, including the number of nodes $N$, edges $E$, the average degree $\langle k\rangle$ and the largest eigenvalue $\lambda_{max}$.} 
\begin{center}
\begin{tabular} {l r r r r r r}
  \hline \hline
   Network     &$N$      &$E$       & $\langle k\rangle$      &$\lambda_{max}$        \\ \hline
   Email       &1133     &5451      &9.60                     & 20.75                 \\
   Protein     &2284     &6646      &5.82                     & 19.04                 \\
\hline \hline
\end{tabular}
\end{center}
\end{table}
\subsection{Measurement}
To evaluate the performance of the SSD method, the Kendall's tau $\tau$ is introduced to measure the correlation of the node spreading influence with SSD method, degree, K-shell and eigenvector centrality. The Kendall's tau $\tau$ is used to measure the correlation between two ranking lists. The Kendall's tau $\tau$ value is between [-1,1], and the increasing values imply the method can identify the node spreading influence more accurately. The Kendall's tau $\tau$ is defined as
\begin{equation}\label{equation4}
\tau=\frac{2}{N(N-1)}\sum_{i<j}{\rm{sgn}}[(y_{i}-y_{j})(z_{i}-z_{j})],
\end{equation}
where $N$ is the number of nodes of a network, $y(i)$ is the node spreading influence of node $i$, $z(i)$ are the values generated by the SSD method, degree, K-shell and eigenvector centrality and sgn$(x)$ is a piecewise function, when $x>0$, sgn$(x)=+1$; $x<0$, sgn$(x)=-1$; when $x=0$, sgn$(x)=0$.
\subsection{Numerical results}
\begin{figure}[ht]
\center\scalebox{0.66}[0.66]{\includegraphics{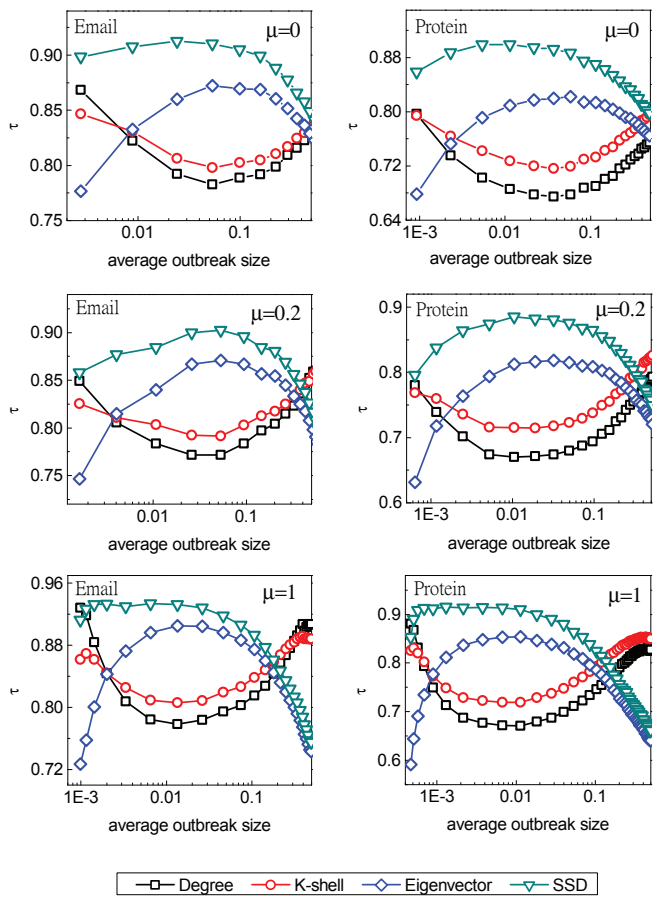}}
\caption{(Color online) The Kendall's tau values $\tau$ obtained by comparing the ranking list generated by the SIR spreading process and the ranking lists generated by the degree (squares), K-shell (circles), eigenvector (diamonds) and SSD method (triangles) with recovering rate $\mu$ 0, 0.2, 1 respectively. The average outbreak size (horizontal axis) being controlled by the spreading rate $\beta$ is the average number of the infected nodes when choosing the initial node of of the network. From which one can find that the SSD method could identify the node spreading influence more accurately than other methods. The results are averaged over 100 independent runs with different spreading rate $\beta$ when the average outbreak size reach 50\% of the network.}\label{Fig2}
\end{figure}
\begin{figure}[ht]
\center\scalebox{0.66}[0.66]{\includegraphics{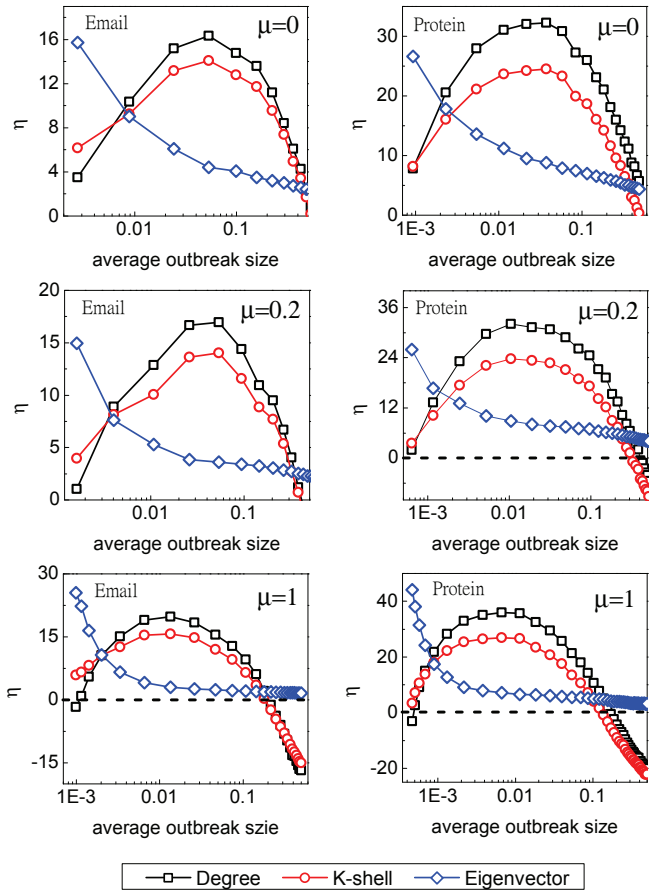}}
\caption{(Color online) The vertical axis $\eta$ is the improved ratio $\eta$ for degree, K-shell and eigenvector centrality with different spreading rate $\beta$ on two real networks. From which one can find that the improved ratio $\eta>0$ indicates the Kendall's tau for SSD method is higher than other Kendall's tau generated by other methods. The results are averaged over 100 independent runs with different spreading rate $\beta$ when the average outbreak size reach 50\% of the network.}\label{Fig3}
\end{figure}
\begin{figure}[ht]
\center\scalebox{0.66}[0.66]{\includegraphics{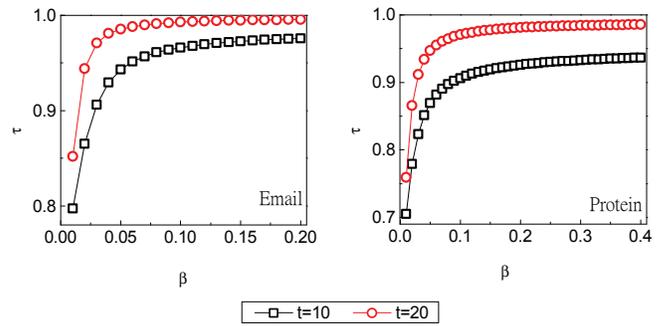}}
\caption{(Color online) The Kendall¡¯s tau values $\tau$ obtained by comparing the ranking list generated by the SSD method and the ranking lists generated by eigenvector centrality when the recovering rate $\mu=0.2$ and time step is 10 (squares) and 20 (circles) for Email and Protein network. From which one can find that the Kendall¡¯s tau $\tau$ of the SSD method and eigenvector centrality is almost equal to 1 when the spreading rate $\beta$ and the time step is large, which indicates the ranking list generated by the SSD method have the same ranking list by the eigenvector centrality for Email and Protein network.}\label{Fig3}
\end{figure}

In this section we check the performance of the SSD method by the Kendall's tau $\tau$. As shown in Fig. 2, the Kendall's tau values $\tau$ of the SSD method is between 0.66 and 0.93, which indicates that the ranking list generated by the SSD method are highly identical to the ranking list by the SIR spreading process. The comparisons between the SIR model and the SSD method show that the nodes with influential neighbors will have larger spreading influence. Comparing with degree, K-shell and eigenvector centrality, the Kendall's tau $\tau$ of the SSD method would be much better than the ones generated by other methods, which indicates that the SSD method can identify the node spreading influence more accurately than degree, K-shell and eigenvector.

Figure 3 reports the improved ratio in the Kendall's tau $\tau$ when applying the SSD method compare with degree, K-shell and eigenvector eigenvector. The improved ratio is defined as
\begin{equation}\label{equation2}
\eta=\frac{\tau^{S}-\tau^{0}}{\tau^0},
\end{equation}
where $\tau^{S}$ is the Kendall's tau of the SSD method, $\tau^{0}$ is the Kendall's tau of degree, K-shell and eigenvector respectively. Clearly, $\eta>0$ indicates an advantage of the SSD method. The improved ratio in $\tau$ for degree, K-shell and eigenvector with different spreading rate $\beta$ and recovering rate $\mu$ on two real networks are shown in Fig. 4. From which one can find that the ranking accuracy has been remarkably improved by the SSD method in different methods. The largest improved ratio $\eta$ for degree, K-shell and eigenvector could reach 35.9\%, 27.0\% and 44.1\% respectively.

However we can find that the kendall'a tau $\tau$ decreases with the increase of the spreading rate $\beta$ when the recovering rate $\mu=1$ for Email and Protein network in Fig. 2 and the improved ratio is even lower than 0 for large spreading rate $\beta$,  which indicates the SSD method fails to identify the node spreading influence with large spreading rate $\beta$. Because SSD method is an approximate method for calculating the node spreading influence and there are two disadvantages in SSD method. Firstly, it does not consider the node state at time step $t-1$ when calculating the probability of the nodes to be infected at time step $t$ by equation (3). Secondly, the SSD method calculates the probability of a node which has two infected nodes to be infected by linear method stead of non-linear method. For example, according to equation (3) if a susceptible node $i$ has two infected neighbour nodes at time step $t-1$, the probability of node $i$ to be infected at time step $t$ is $2\beta$ instead of $1-(1-\beta)^2$.

We can find that the curve of SSD method has the same trend with eigenvector centrality. Especially the Kendall's tau $\tau$ of the SSD method is the same as eigenvector method with large spreading rate $\beta$. Figure 4 reports the correlation between the SSD method and the eigenvector centrality with different spreading rate $\beta$ and time step $t$ when the recovering rate $\mu=0.2$. From which one can find that the Kenall's tau $\tau$ of the ranking list generated by SSD method and eigenvector method increases with the spreading rate $\beta$, which indicates the ranking list generated by the SSD method is the same as the one generated by eigenvector method with large spreading rate $\beta$ and time step $t$ which is proved in the section 3.
\section{conclusion}
In this paper, we propose a general framework for identifying the node spreading influence by combining the network structure and the spreading dynamics. By theoretical analyzing the spreading differential equation, one can get that the total number of the infected node for one target node is determined by the adjacent matrix $A$, spreading parameter $\beta$ and initial state of the target node. Therefore, we propose a structure spreading dynamics (SSD) method for ranking the node spreading influence. The simulation results for two real networks show that the Kendall's tau $\tau$ of the SSD method is between 0.66 and 0.93, which indicates that the ranking list generated by the SSD method is highly identical to the ranking list by the SIR spreading process. Comparing with the degree, K-shell and eigenvector centrality, the largest improved ratio $\eta$ could reach 35.9\%, 27.0\% and 44.1\% respectively. Furthermore we can find that the ranking list generated by the SSD method is almost the same as the one generated by eigenvector centrality with large spreading rate $\beta$ and time step $t$ as we analyze.

However, the kendall'a tau $\tau$ of the SSD method decreases with the increase the spreading rate $\beta$ when the recovering rate $\mu=1$ in Email and Protein network, which indicates the SSD method could not identify the node spreading influence very well for large spreading rate $\beta$. Because SSD method is an approximate method and there are two disadvantages in this method. Firstly, it does not consider the node state at time step $t-1$ when calculating the probability of the nodes to be infected at time step $t$ by equation (3). Secondly, the SSD method calculates the probability of a node which has two infected neighbour nodes to be infected by linear method instead of non-linear method. The solving of the above problems can help us to improve the accuracy of the SSD method for identifying the node spreading influence and study the multiple-nodes spreading process.

\begin{acknowledgments}
The authors wish to thank Dr. Tao Zhou for discussion. This work is supported by the National Natural Science Foundation of China (Nos. 71171136), the Shanghai Leading Academic Discipline Project of China (No. XTKX2012), MOE Project of Humanities and Social Science (No. 13YJA630023), the Foundation of Shanghai Research Institute of Publishing and Media (No. SAYB1407).
\end{acknowledgments}

\end{document}